\documentclass[preprint,prb,floats,superscriptaddress]{revtex4}
\usepackage{amsmath}
\usepackage{graphicx}
\newcommand{\be}{\begin{equation}} 
\newcommand{\ee}{\end{equation}}
\newcommand{\bea}{\begin{eqnarray}}
\newcommand{\eea}{\end{eqnarray}}

\newcommand{\rr}{{\bf r}}
\newcommand{\NN}{{\bf \nabla}}
\newcommand{\FF}{{\bf F}}

\newcommand{\vv}{{\bf v}}

\newcommand{\uu}{{\bf u}}

\renewcommand{\ss}{{\bf\sigma}}

\newcommand{\shat} {\hat{\ss}}

\begin{document}
\date{\today}
\title{A unified framework for dynamic density functional and Lattice Boltzmann methods}
\author{Umberto Marini Bettolo Marconi}
\address{Dipartimento di Fisica, CNISM and INFN, Via Madonna delle Carceri,
68032 Camerino (MC), Italy}
\author{Simone Melchionna}
\address{INFM-CNR and Dipartimento di Fisica, Universit\`a La Sapienza,
00185 Rome, Italy}

\begin{abstract}
  
Using methods of kinetic theory and liquid state theory we propose a description of the non-equilibrium behavior 
of molecular fluids which takes into account their microscopic structure and thermodynamic properties.
The present work represents an alternative to the recent
dynamic density functional theory which can only deal with colloidal fluids
and is not apt to describe the hydrodynamic behavior of a  molecular fluid. 
The method is based on a suitable modification of the Boltzmann transport equation for
the phase space distribution and provides
a detailed description of the local structure of the fluid and of the transport coefficients under inhomogeneous conditions.
Finally, we  propose a practical scheme to solve numerically and efficiently the resulting kinetic equation by
employing a discretization procedure analogous to the one used in the Lattice Boltzmann method.

\end{abstract}

\maketitle
PACS (47.11.-j,47.61.-k,61.20.-p)
\section{Introduction}

Understanding the dynamics of fluids under non-equilibrium conditions
is of capital importance for fundamental statistical physics as well
as applied disciplines such as engineering, fluid mechanics, rheology,
and physiology.
In recent years there has been an upsurge of interest towards the study of
transport phenomena in strongly inhomogeneous systems mostly motivated
by important physical and technological
applications such as microfluidics\cite{Squires}, 
colloids, oil recovery, lab-on-a-chip  devices etc.
These examples require the knowledge of structural
and dynamical fluid properties in the presence of restricted geometries
and/or structured substrates and of
external gradients or time dependent fields. 
Typical time and length scales involved
can be significantly shorter than those usually assumed in
standard thermodynamic and hydrodynamic theories.
 In order to 
go beyond these descriptions, several theoretical approaches have been developed
which differ not only by the nature of
systems under scrutiny (colloidal systems have different dynamical behaviors
in comparison with simple fluids), but also from subjective factors
such as the individual scientific background and personal taste. 
These methods include
the dynamic density functional approach, the kinetic approach, mesoscopic methods and
methods based on effective free energies. 
In the present paper  we shall be concerned in some detail only with the first two.

In the last thirty years,
massive efforts have been devoted to develop techniques to study the properties
of non-uniform interacting many particle systems, 
among these Density Functional theory (DFT) 
being perhaps the most versatile \cite{Evans1,Evans2}. In DFT the single-particle density
profile $\rho(\rr)$ is obtained via minimization of the  Grand potential
functional. Functional derivatives of the Grand potential determine
the multi-particle correlation functions and 
all the structural and thermodynamic
properties for a system with arbitrary inhomogeneity.
Recently,  dynamical generalizations 
of these equilibrium methods have been applied to non-equilibrium 
problems such as diffusion, Stokes drift, polymeric fluids confined 
to cavities, etc.
The dynamical density functional ( DDF) is apt to describe the relaxation of Brownian 
particles in a medium and can  be applied in situations where the local
velocity distribution is very close to the Maxwellian  and the density
field varies slowly in time \cite{Tarazona,ArcherEvans}. 
The method is based on the
assumption of instantaneous equilibrium, i.e. that the correlation
functions at a given instant are identical to those of the same 
equilibrium system having the same equilibrium density profile.
In the case of overdamped dynamics, which characterizes colloidal 
particles immersed in a solvent,
the evolution is mainly governed by structural rearrangements
so that such an interplay is evident and is justified to describe the
evolution of the system in terms only of its density 
and density correlations. It is therefore not surprising that in DDF the velocity field is 
slaved to the density field and does not play any autonomous role.

On the other hand, in molecular fluids the momenta of the particles are not damped
by the interaction with the solvent so that the total momentum 
is  conserved and the momentum current must be 
treated on equal footing as the local density \cite{Espanol,Archer2005}. 
Hydrodynamics describes the non-equilibrium state of a system
by means of a set of local variables which are density, 
momentum and temperature of the fluid \cite{DeGroot}.
However, hydrodynamics does not apply to phenomena
which are not slowly varying in space and one has to
consider a finer level of description, such as one based on 
a suitable generalization of the Boltzmann transport equation for $f(\rr,\vv,t) $,  
the phase space density distribution
of particles with position $\rr$ and velocity $\vv$ at time $t$.
  The modeling of the interactions in the transport equation
depends on the nature of the fluid and on the degree of accuracy required.
The widely used Boltzmann collision operator has been studied 
for several kinds of interaction potentials
and gives predictions for the transport coefficients, but does not provide
an accurate representation of the thermodynamics and structural properties of fluids 
\cite{Harris,Bird,Kreuzer,McQuarrie}.
For simple applications one can even approximate further the interaction term
by a linear relaxation term, as proposed  by Bhatnagar,Gross and Krook (BGK)
\cite{BGK}. 
A more refined approximation
is needed if one is interested in 
a dynamical description taking into account
both the equilibrium properties of the fluid, 
such as the equation of state and the equilibrium
density profile in an external force field, and the transport 
properties under inhomogeneous conditions. The Revised Enskog theory (RET), 
originally introduced by Van Beijeren and Ernst for hard sphere systems \cite {VanBeijeren}, has the ability to describe 
the local fluid structure within a kinetic formalism.
It can serve as a reference theory to study fluids with 
different types of interactions.
However, even in the RET case a numerical solution is often 
too demanding in terms of computer speed and memory so that one has to resort to 
some simplifications. One of these has been proposed more than a decade ago
by Dufty, Santos and Brey \cite{Brey} and consists in separating the slowly evolving part
of  $f(\rr,\vv,t)$, associated with
the five hydrodynamic modes, from the fast non-hydrodynamic modes. 
 
In the present work, we will show how to obtain a workable numerical method based on this equation
which includes information about the microscopic nature of the fluid and contains hydrodynamics as a limiting case.
This paper is organized as follows: In section \ref{generalsection} we introduce the transport equation and 
the collision operator, we define the hydrodynamic fields and derive the balance equations. In section \ref{colloid}
we briefly derive, for the sake of comparison,  the DDF equation, in section \ref{molecular}
we derive the equation of evolution for molecular fluids which differs from the DDF equation
because it considers the evolution of the density, velocity and temperature fields altogether.
In Sec. \ref{lbesolution}, we propose to use the Lattice Boltzmann method as a strategy to solve numerically
the transport equation. Finally, in Sec. \ref{conclusions} we summarize and draw some conclusions.

\section{Kinetic Theory}
\label{generalsection}
We consider a simple fluid, whose elementary constituents, the molecules, 
mutually interact via a pairwise, spherically symmetric potential $U(r-r')$.
The statistical description of such a system is based on the exact
BBGKY hierarchy of dynamical equations for the reduced distribution functions, 
whose first level is the following kinetic equation

\begin{eqnarray}
\partial_{t}f(\rr,\vv,t) +\vv\cdot\NN f(\rr,\vv,t)
+\frac{\FF(\rr)}{m}\cdot\frac{\partial}{\partial \vv} f(\rr,\vv,t)
= \Omega(f|\rr,\vv,t)+B(f|\rr,\vv,t)
\label{uno}
\end{eqnarray}
where $f(\rr,\vv,t)$ is the one particle phase space density distribution
at time $t$ and at the point $(\rr,\vv)$,
$\FF(\rr)$ is an external velocity-independent force field,
$\Omega(f|\rr,\vv,t)$  represents the effect on the single particle distribution function of the
interactions among fluid particles
and $B(f|\rr,\vv,t)$ is a coupling 
to an external agent, usually termed heat-bath. 
The interaction term
is given by the following exact expression
\be
\Omega(\rr,\vv,t)=
\frac{1}{m}{\NN_{\bf v}} \int d\rr'\int d\vv' 
f_2(\rr,\vv,\rr',\vv',t){\bf \NN_{\bf r}}U(|\rr-\rr'|) 
\label{Ke} 
\ee
 involving the two particles distribution function,
 $f_2(\rr,\vv,\rr',\vv',t)$, 
which in turn depends on the three particle correlation function.
However,
in the simplest closure schemes $\Omega(f)$ (and $B(f)$) 
can be expressed in terms 
only of $f(\rr,\vv,t)$ so that \eqref{uno} becomes self-consistent
and one can devise practical schemes of solution.

In order to avoid the difficult calculation of the  
two-particle distribution function, one can introduce
an approximate truncation of the BBGKY hierarchy. That is,
we assume the following factorization of $f_2$
\be
f_2(\rr,\vv,\rr',\vv',t)= 
f(\rr,\vv,t) f(\rr',\vv',t) g_2(\rr,\rr',t|n)
\label{factor}
\ee
where $g_2$ is the equilibrium static pair correlation function, a functional of the local density.

One can further coarse grain the description and
use the kinetic equation \eqref{uno} to derive, by the usual
Chapman-Enskog method, the macroscopic hydrodynamic equations,
where a limited set, ($d+2$), fields, namely
density, momentum current and energy density are assumed to 
represent the state of the fluid. To achieve this goal and
obtain an autonomous set of equations by relating the currents of the
hydrodynamics fields to the gradients of these fields,
one needs to find first the form of the distribution $f(\rr,\vv,t)$
when the system is perturbed  from equilibrium.
Having the perturbed form of $f(\rr,\vv,t)$, one can
compute the transport coefficients and thus relate currents
and fields, without assuming the phenomenological constitutive 
relations. The transport coefficients
are usually studied under conditions of almost uniform
density, so that the spatial dependence of $f(\rr,\vv,t)$ can be 
taken into account only at linear order.
However, under many circumstances 
the presence of density gradients
strongly influences the thermodynamic and dynamical properties
of a fluid so that it seems important to study the interplay
between structure and dynamics.

\subsection{Hydrodynamic fields and balance equations}

A direct solution of eq. \eqref{uno} in terms of the unknown $f(\rr,\vv,t)$
is clearly beyond reach. 
We thus follow a different strategy, originally proposed by Dufty, Santos and 
Brey in a study of the
hard-sphere system. 
Their approximation is tantamount to treat separately
and accurately  the part of $\Omega(f)$ 
contributing to the evolution equations for the hydrodynamic
fields from the part relative to the non-hydrodynamic 
moments of the distribution,
which can be safely approximated by a simple relaxation time ansatz.

A convenient way to analyze eq. \eqref{uno} is to consider
the equation for the lowest velocity moments of the
distribution function,
which correspond to the standard five hydrodynamic fields
describing the slowly evolving part of $f(\rr,\vv,t)$.
We first introduce the 
average local density
\be
n(\rr,t)= \int d\vv f(\rr,\vv,t),
\label{density}
\ee
the average local fluid velocity
\be
\uu(\rr,t)=\frac{1}{n(\rr,t)}\int d\vv \vv f(\rr,\vv,t)
\label{momentum}
\ee
and the average local temperature
\be
T(\rr,t)=
\frac{m}{3k_B n(\rr,t)}\int d\vv (\vv-\uu)^2 f(\rr,\vv,t)
\label{energy}
\ee
where $k_B$ is the Boltzmann constant.
We shall employ throughout the paper 
the Einstein summation convention
that repeated indices are implicitly summed over and the notation 
$\partial_i$ to indicate the partial derivative w.r.t. the 
$i-th$ cartesian component of the vector $\rr$ and $\partial_t$ to indicate
the partial derivative w.r.t. time.

A  set of balance equations are
obtained for density and momentum 
by multiplying both sides of eq. \eqref{uno}
by $1,$ and $\vv$, respectively,  and integrating w.r.t. velocity 
\be
\partial_{t}n(\rr,t) +\partial_i ( n(\rr,t) u_i(\rr,t))=0
\label{continuity}
\ee 
and
\be
m n(\rr,t)[\partial_{t}u_j(\rr,t)+u_i(\rr,t) \partial_i u_j(\rr,t)] 
+\partial_i P_{ij}^{(K)}(\rr,t) -F_j(\rr)n(\rr,t)- C_j^{(1)}(\rr,t)=
b_j^{(1)}(\rr,t)
\label{momentumcontinuity}
\ee

An analogous  balance equation for the temperature field can be derived 
by multiplying  
\eqref{uno} by $m(\vv-\uu)^2/2$ and integrating w.r.t. $\vv$
\be
\frac{3}{2}n(\rr,t) [\partial_{t} T(\rr,t)+u_i(\rr,t)\partial_i T(\rr,t)]+P^{(K)}_{ij}(\rr,t) \partial_i u_j(\rr,t)
+\partial_i q^{(K)}_i(\rr,t)-C^{(2)} (\rr,t) =b^{(2)}(\rr,t)
\label{energyequation}
\ee

To establish the momentum and temperature
equations \eqref{momentumcontinuity} and \eqref{energyequation}
we introduced some 
quantities, which in general cannot be 
expressed in terms of the hydrodynamic fields. These are the 
kinetic components of the pressure
tensor, indicated with the superscript "K",
\be
P^{(K)}_{ij}(\rr,t)=m\int d\vv f(\rr,\vv,t)(v-u)_i(v-u)_j
\label{pkin}
\ee
and  the kinetic components of the  heat flux  vector
\be
q^{(K)}_{i}(\rr,t)=\frac{m}{2}\int d\vv f(\rr,\vv,t)(\vv-\uu)^2(v-u)_i
\label{qkin}
\ee
In addition, we defined two terms
stemming from molecular interactions (recall that $\int \Omega d\vv=0$ and $\int B d\vv=0$, because the number of particles is conserved) 
\be
C_i^{(1)}(\rr,t)  =m\int d\vv(v-u)_i  \Omega(f|\rr,\vv,t).
\label{divpressurea}
\ee
and 
\be
C^{(2)} (\rr,t) =\frac{m}{2}
\int d\vv   (\vv-\uu)^2\Omega(f|\rr,\vv,t)
\label{divqc}
\ee
and two terms stemming from the coupling with  the heat bath
\be
b_i^{(1)}(\rr,t)  =m\int d\vv(v-u)_i  B(f|\rr,\vv,t).
\label{divpressureaa}
\ee
and 
\be
b^{(2)} (\rr,t) =\frac{m}{2}
\int d\vv   (\vv-\uu)^2 B(f|\rr,\vv,t)
\label{b2}
\ee

Following ref. \cite{Kreuzer} one can relate the term $C_i^{(1)}$ with
the excess part of  the 
pressure tensor , $P_{ij}^{(C)}(\rr,t)$ ,  
stemming from the interactions,
\be
C_i^{(1)}(\rr,t)=-\partial_j  P_{ij}^{(C)}(\rr,t),
\label{divpressureab}
\ee
where
the components of the excess part (over the ideal gas) of the pressure tensor 
can be computed using a method due to 
Irving and Kirkwood which gives the  following exact  expression
\be
P_{ij}^{(C)}(\rr,t)=-\frac{1}{2}\int_0^1 d \lambda
\int d \rr_{12}\int d\vv d\vv'\frac{\rr_{12}\rr_{12}}{|\rr_{12}|}
\frac{d U(|\rr_{12}|)}{d|\rr_{12}|}
f_2(\rr+(1-\lambda)\rr_{12},\vv,\rr-\lambda\rr_{12},\vv',t)
\ee
with $\rr_{12}=(\rr-\rr')$.
With such an identification, eq. \eqref{momentumcontinuity} assumes the form 
of the standard macroscopic equation for momentum balance.
In general, no analogous relation  exists between  $C^{(2)}(\rr,t)$  and the excess component of the
heat flux, and one should instead take into account also the transfer 
of potential energy in order to derive
an expression in terms of macroscopic fluxes.


\section{Colloidal dynamics and density functional strategy}
\label{colloid}
Let us go back to the non-equilibrium case.
Equation \eqref{uno} contains as limiting cases the Hamiltonian dynamics
and the fully underdamped dynamics when the heat bath
exerts a large friction on the particles.  
In the standard description of the behavior of
colloids it is assumed that the particles are 
subject to two kinds of forces from the surrounding fluid
1) a deterministic friction, proportional
to their velocity and described by Stokes law and 2) a stochastic
force, characterized by a white noise spectrum, 
resulting from the interactions with the solvent molecules.
In formulas, each particle is subjected to the following 
combination of solvent forces $-m\gamma \vv +\sqrt{2\gamma m k_B T}\xi $, 
where $\xi$ is a Gaussian white noise and $\gamma$ a friction coefficient.
A fluctuation-dissipation relation has been assumed between 
the noise amplitude 
and the friction coefficient so that the 
steady state velocity distribution is Maxwellian \cite{Cecconi2005}. 
The effect of these forces  
on the particle phase space distribution $f(\rr,\vv,t)$
can be represented by the following term in eq. \eqref{uno} \cite{Risken} 
\be
B^{(coll)}(f|\rr,\vv,t)=\gamma[\frac{k_B T}{m}\frac{\partial^2} 
{\partial \vv^2} 
+\frac{\partial} {\partial \vv}\cdot \vv ]f(\rr,\vv,t)
\label{colloidheatbath}
\ee
Using the specific form of the heat-bath term, eq. \eqref{colloidheatbath},
we shall now recover the dynamic density functional equation.
In this case $b_i^{(1)}(\rr,t)=-m\gamma n(\rr,t) u_i(\rr,t)$.
The reduction of eq. \eqref{uno} to a DDF dynamics
can be done systematically by employing a multiple time scale approach, as
used by Marconi and Tarazona \cite{Marconi2006,Marconi2007,Melchionna2007}.
The method eliminates systematically all high moments of the velocity
distribution function except the zeroth moment, that is the number density,
by employing an asymptotic expansion in the inverse
friction parameter $\gamma^{-1}$.
The result is an equation for
the density field containing corrections to the standard non-linear diffusion 
equation usually employed in DDF.
However, for the present scope, we shall employ an heuristic 
and much simpler method to contract the phase space description 
of eq. \eqref{uno} into the DDF description.

 We start by considering an approximate solution of the transport equation of the form
 \be
f_{DDF}(\rr,\vv,t)=n(\rr,t)[1+\frac{\vv\cdot\uu(\rr,t)}{v_T^2}]\phi_0(\vv)
\label{trialddf}
\ee
with $\phi_0(\vv)=[\frac{1}{2 \pi v_T}]^{3/2}\exp
\Bigl(-\frac{m\vv^2}{2 v_T^2}\Bigl)  $ and $v_T^2=k_B T/m$.
Substituting eq. \eqref{trialddf} into \eqref{uno}
\bea
&&\phi_0(\vv) \{[\partial_t n(\rr,t)+\partial_i (n(\rr,t) u_i(\rr,t)]+
\nonumber\\
&&\frac {v_i}{v_T^2}
[\partial_t (n(\rr,t)u_i(\rr,t))+v_T^2 \partial_i n(\rr,t) -\frac{F_i}{m} 
n(\rr,t)+\gamma n(\rr,t) u_i(\rr,t)]+
\nonumber\\
&&(\frac{v_j v_j}{v_T^2}-\delta_{ij}) [\partial_i (u_j(\rr,t) n(\rr,t))-\frac{F_i}{m}u_i(\rr,t) n(\rr,t)]=\Omega(\rr,\vv,t)
\label{array}
\eea
Multiplying by $1$ and by $v_i$ eq.\eqref{array}
and integrating w.r.t. velocity, we obtain the continuity equation
\eqref{continuity}
and the following equation for the momentum current
\be
m \partial_{t}(n(\rr,t) u_i(\rr,t)) +\gamma m n(\rr,t) u_j(\rr,t)
+k_B T \partial_i n(\rr,t)(\rr,t) -F_i(\rr)n(\rr,t)- C_i^{(1)}(\rr,t)=
0,
\label{momentumddft}
\ee
The DDF equation is recovered if one takes $\gamma$ large and 
neglects the time derivative
of the momentum current,  
$n(\rr,t) u_i(\rr,t))$,
viz. one neglects the time derivative in the l.h.s of eq. 
\eqref{momentumddft}.
Using the continuity equation \eqref{continuity} 
to eliminate the current one obtains
the following diffusion equation   
\be
 \partial_{t}n(\rr,t) =\frac{1}{m \gamma}
\partial_i [k_B T \partial_i n(\rr,t) -F_i(\rr)n(\rr,t)- C_i^{(1)}(\rr,t)]
\label{ddfteq}
\ee
In order to identify the full DDF equation one must specify the form of the
collision term $C_i^{(1)}(\rr,t)$.

We begin by a simple ansatz for the interaction term 
\eqref{Ke} by assuming the following adiabatic approximation
\be
\Omega_{DDF}(\rr,\vv,t)=
\frac{1}{m}\frac{\partial}{ \partial v_i} \int d\rr'\int d\vv' 
f(\rr,\vv,t) f(\rr',\vv',t) g_2(\rr,\rr',t|n) \frac{\partial}{\partial r_i} U(|\rr-\rr'|) 
\label{Ke2} 
\ee
where we have approximated the two-particle correlations   
by those of an equilibrium system having  the same density profile
as the system at time $t$. 
Next, we define the molecular field as
\be
F_i^{(mol)}(\rr,t)=-\int d\rr' n(\rr',t) g_2(\rr,\rr',t|n)\frac{\partial } {\partial r_i} U(|\rr-\rr'|) 
\ee
so that the interaction  term becomes

\be
\Omega_{DDF}(\rr,\vv,t)=
-\frac{1}{m}\frac{\partial}{ \partial v_i}f(\rr,\vv,t)F_i^{(mol)}(\rr,t)
\label{Ke3} 
\ee
By computing the corresponding $C_i^{(1)}$ from \eqref{divpressurea}
the result is
\be
C_i^{(1)}(\rr,t) =n(\rr,t) F_i^{(mol)}(\rr,t).
\label{c1}
\ee

It is straightforward to show that expression \eqref{c1} can be recast as
\be
C_i^{(1)}(\rr,t) =-n(\rr,t)\frac{\partial } {\partial r_i} \frac{\delta\Delta  {\cal  F}} {\delta n(\rr,t)} 
\ee 
where $\Delta {\cal  F}$ is the free energy excess over the ideal gas \cite{Tarazona} .

Notice that,
in the case treated above, both the noise and the friction are externally
imposed by the presence of the solvent and such an heat bath term breaks the translational invariance of the
system, since the heat bath is assumed to be at rest.

Unfortunately, as shown in ref. \cite{Marconi2006}, the method cannot be extended
to systems with small  or vanishing friction proportional to the velocity  and therefore  cannot
be extended to  describe dense liquids, which are instead
characterized by an internal friction proportional to velocity gradients and not to the velocity itself.
The slaving of the momentum current and of the energy current to the density, which allows
the reduction of the complex dynamics eq. \eqref{uno} to eq.  \eqref{ddfteq} is at work only in the high friction regime.
Hence, a different approach is needed to treat molecular fluids.

\section{Molecular fluids and modified Boltzmann equation strategy}
\label{molecular}

The dynamics of molecular fluids is described by Newton's equation of motion,
as opposed to colloidal fluids where the presence of the solvent 
corresponds to overdamped Brownian dynamics
encapsulated in the DDF equation.
Two important properties of molecular fluids 
must be preserved in any theoretical representation. These are
the Galilei invariance, with respect to the
reference frames moving relative to each other at constant velocity, 
and the momentum conservation.
These are relevant features of hydrodynamics and are not displayed by the 
DDF method, where the absolute velocities are damped, thus 
privileging the ``solvent reference frame''. In the case of molecular fluids,
the total momentum must be conserved, so that a heat bath
such as that used in the previous section is not acceptable.
In a molecular fluid the dissipation is determined by
other mechanisms such as, among the others,  
the internal friction proportional to 
velocity gradients (viscosity), and not to velocity itself, and to
temperature gradients (thermal conductivity).
We shall use this information in order to make approximations 
not violating these conservation laws, which are at the origin of the low 
frequency hydrodynamic modes.

We start by splitting $\Omega$ into two
contributions, the first taking into account  accurately the effect
of the interaction term on the hydrodynamic variables, so that
a correct thermodynamic and structure
of the fluid is achieved, and the second describing in lesser detail 
the evolution of the non-hydrodynamic modes.

Based on the form of eqs.\eqref{density}, \eqref{momentum}
and \eqref{energy}, we write
\be
\Omega(f|\rr,\vv,t)=\frac{ \phi_M(\rr,\vv,t)}{k_B T(\rr,t)}
\Bigl((\vv-\uu)\cdot
{\bf C}^{(1)}(\rr,t)
+(\frac{ m(\vv-\uu)^2}{3 k_B T(\rr,t)} -1 )C^{(2)}(\rr,t)\Bigl)+
\delta\Omega(f|\rr,\vv,t))
\label{brey}
\ee
with
\be
\phi_M(\rr,\vv,t)=[\frac{m}{2\pi k_B T(\rr,t)}]^{3/2}\exp
\Bigl(-\frac{m(\vv-\uu)^2}{2 k_B T(\rr,t)} \Bigl).
\label{localmaxwell}
\ee

Equation \eqref{brey}
represents a separation of the interaction term $\Omega(f)$
into a direct contribution 
to the hydrodynamic modes,
plus a correction, $\delta \Omega$, which is assumed 
not to act directly on these modes.
It is easy to see that the following approximation for the term
$\delta \Omega$ fulfills such a  condition
\be
\delta\Omega(f|\rr,\vv,t)\simeq\nu_0 [\frac{k_B T(\rr,t)}{m}
\frac{\partial^2} {\partial \vv^2} 
+\frac{\partial} {\partial \vv}\cdot (\vv-\uu(\rr,t)) ]f(\rr,\vv,t)
\label{liquidheatbath}
\ee
In fact, by direct inspection one sees that
the form of  $\Omega(f)$  eq.\eqref{brey} together with 
eq.\eqref{liquidheatbath}
reproduces eqs. \eqref{divpressurea} and  \eqref{divqc}, and that 
 $\delta \Omega$ does not contribute  to these two integrals.
The term $\delta\Omega $ is assumed to be linear with respect to $f$
and is approximated by an heat-bath operator. At variance with the
heat bath operator introduced in the colloidal problem, its effect
is to induce a matching of the velocity of the particles to the local value
of the average velocity of the fluid and a matching of the temperature of the
particles to the local value of the average temperature. The coefficient 
$\nu_0$ is a phenomenological adjustable parameter chosen to 
reproduce the value of the viscosity of the fluid. 
The simplified and approximated version of \eqref{uno} 
represented by eq. \eqref{brey} with eq. \eqref{liquidheatbath}
reproduces by construction equations \eqref{continuity}, 
\eqref{momentumcontinuity} and
\eqref{energyequation}. 
We comment that
the form  \eqref{liquidheatbath} preserves the Galilei invariance
of the fluid. This heat bath ``co-moving'' with the fluid
is of course very different from the one introduced in the study of colloids.
Physically it can be thought to  represent the effect of the fast modes on the slow modes,
and can be seen as an intrinsic heat bath, as opposed to the extrinsic heat bath employed in the
study of colloids.

We remark that the invariance of the heat bath with respect to the choice of the
reference frame is the same requirement which has lead to the formulation of the so-called
Dissipative Particle Dynamics (DPD) \cite{HoogerbruggeKoelman, EspanolWarren}. There 
on a phenomenological basis one 
introduces  pairwise dissipative and random forces, a
 ``DPD thermostat'' which locally  conserves the momentum and leads the emergence of hydrodynamic flow effects on the macroscopic scale \cite{Dunweg}. 
The present method instead of introducing a pairwise friction proportional to the velocity difference of two colliding particles, employs
a frictional force proportional to the difference between the individual particle velocity and the average fluid velocity $\uu$.

It is useful to represent $f(\rr,\vv,t)$ as the sum
of a local thermodynamic equilibrium state, $f_{loc}(\rr,\vv,t)=
n(\rr,t)\phi_M(\rr,\vv,t)$, 
plus a contribution representing the deviation from such a state
\be
f(\rr,\vv,t)=f_{loc}(\rr,\vv,t)+\delta f(\rr,\vv,t)
\label{localequi}
\ee
The function
$f_{loc}(\rr,\vv,t)$ alone fully determines the
values of the hydrodynamic moments \eqref{density}, \eqref{momentum}
and \eqref{energy}, whereas
 $\delta f(\rr,\vv,t)$ does not contribute.
On the other hand, $\delta f$ does
contribute to the heat flux and to the viscous pressure tensor as
shown below.  
The interplay between $\delta f$ and the hydrodynamic fields occurs
both via the terms $C^{(1)}$ and $C^{(2)}$, which are functionals of $\delta f$ and $f_{loc}$
and via kinetic components of the pressure tensor and heat flux
vector.

It is relevant to remark that $\delta \Omega$ vanishes when
$f=f_{loc}$. The mathematical properties of the differential operator
featuring in eq. \eqref{liquidheatbath} are well-known, in
particular displaying a non-positive spectrum with discrete eigenvalues
$0,-\nu_0,-2\nu_0,...$.  The local Maxwellian \eqref{localmaxwell}
represents the eigenfunction associated with the null eigenvalue,
whereas the higher-order eigenfunctions are associated with the non
hydrodynamic modes.  However, for practical purposes it is more
convenient to simplify further this term and resort to a
drastic assumption, by replacing the heat bath term \eqref{liquidheatbath}
by a BGK-like relaxation term
\be
\delta \Omega (f|\rr,\vv,t)=
-\nu_0 \Bigl( f(\rr,\vv,t)-n(\rr,t)\phi_M(\rr,\vv,t)\Bigl)=
-\nu_0 \delta f(\rr,\vv,t)
\label{simple}
\ee
which keeps the relevant properties of \eqref{liquidheatbath}, at the price of assuming a single relaxation time,
$\nu_0^{-1}$, for all non-hydrodynamic modes.

Here, $\nu_0$ is a 
phenomenological parameter, representing a 
collision frequency, chosen as to reproduce the 
viscosity of the fluid. 
The above approximation replaces by a much simpler relaxation  the complicated 
effects of the interactions among the non-hydrodynamic moments.
The action of $\delta \Omega$ 
is to induce a rather fast relaxation of
the solution towards a state of local equilibrium \cite{marra_lebowitz}.
Once such a state has been reached the system evolves
towards the steady state via equilibration of different
regions through exchange of hydrodynamic fluxes.

In addition, since also the term enforcing the local equilibrium condition \eqref{liquidheatbath}
contains as local parameters $\uu(\rr,t)$ and $T(\rr,t)$, the system has the correct
long wavelength properties required by hydrodynamics.
For future reference we rewrite the combination of  eq.\eqref{uno}, \eqref{brey} and \eqref{simple}   
\begin{eqnarray}
&&\partial_{t}f(\rr,\vv,t) +\vv\cdot\NN f(\rr,\vv,t)
+\frac{\FF^{ext}(\rr)}{m}\cdot\frac{\partial}{\partial \vv} f(\rr,\vv,t)-
\nonumber\\ &&
\frac{ \phi_M(\rr,\vv,t)}{k_B T(\rr,t)}
\Bigl((\vv-\uu)\cdot
{\bf C}^{(1)}(\rr,t)
+(\frac{ m(\vv-\uu)^2}{3 k_B T(\rr,t)} -1 )C^{(2)}(\rr,t)\Bigl)=
-\nu_0\delta f\rr,\vv,t)
\label{unobrey}
\end{eqnarray}
The advantage of using eq. \eqref{unobrey} instead 
of the apparently equivalent set of coupled hydrodynamic
equations is the following.
Even in the non-interacting case the moments of the distribution function are coupled 
and to interrupt the hierarchy one needs a truncation scheme. This truncation can be avoided by working directly
with  eq. \eqref{unobrey}, where one can apply
the powerful technique of the Lattice Boltzmann equation (LBE). In the LBE
one discretizes the velocity and the coordinates of the
particles using a finite grid and computes directly the distribution function
$f(\rr,\vv,t)$. This program would had been numerically too demanding if
we had to use eq. \eqref{uno} because of the large number of integrals involved
in the evaluation of the interaction term.

Eq. \eqref{unobrey} treats the hydrodynamic moments
in a privileged fashion and describes a rapid 
equilibration of the system toward the
local equilibrium state. The remaining stage is described  by
hydrodynamics, that is by mass, momentum and energy transport on larger scales.

\subsection{Approximate solution, kinetic contribution to the viscosity and heat conduction}

Although this procedure is not
necessary in numerical work, in order to gain some insight, 
in the present section we shall determine 
$\delta f$ perturbatively starting from $f_{loc}$, the local equilibrium
state. We
substitute $f\simeq  f_{loc}$ in  the l.h.s. of \eqref{unobrey}  
and obtain an equation for $\delta f$ in terms of derivatives 
of the hydrodynamic fields.

The details of the calculation can be found in standard textbooks \cite{Huang} and hence will be skipped here. 
The substitution of $f_{loc}$ in the left hand side of eq.  \eqref{unobrey} gives
\bea 
&&\Bigl \{ \bigl [ \partial_t n+\partial_i(n u_i) \bigr ]+ \bigl
[ n \partial_t u_k+n u_i \partial_i(n u_k)+\partial_k (n
T)-\frac{F_k}{m}-C^{(1)}_k\bigr ]\frac{m(v_k-u_k)}{k_BT} \nonumber\\
&& +\frac{n}{2T }\bigl [ \partial_t T+u_i \partial_i T +\frac{2}{3}
T\partial_i u_i -\frac{2}{3n} C^{(2)}\bigr ] (\frac{m(\vv-\uu)^2}{k_B
T}-3) +\frac{n}{T }\partial_i T \bigl (m\frac{(\vv-\uu)^2}{2
k_BT}-\frac{5}{2} \bigr)(v_i-u_i) \nonumber\\ 
&& +m \frac{n}{k_B
T}\Bigl ( (v_i-u_i)(v_k-u_k)- \frac{(\vv-\uu)^2}{3}\delta_{ij}
\Bigr)\partial_i u_k \Bigr\}\phi_{M}(\rr,\vv,t)=-\frac{\delta
f(\rr,\vv,t)}{\nu_0} 
\eea

Since $\delta f$ does not contain terms proportional to the first
three terms of the above equation, because it must have vanishing
hydrodynamic moments, we must impose that the first three terms in the
l.h.s. vanish. These are the so-called solvability conditions and are
precisely the balance equations \eqref{continuity},
\eqref{momentumcontinuity} and \eqref{energyequation} at the Euler
level, i.e. without kinetic viscosity and heat conduction
contributions.

We thus obtain the following explicit representation of $\delta f$
\bea \delta f&=&-\frac{1}{\nu_0}\frac{n(\rr,t)}{T} \phi_M(\rr,\vv,t)
\Bigl[ \Bigl ( \frac{m}{2}\frac{(\vv-\uu)^2}{k_B T(\rr,t)} -
\frac{5}{2} \Bigr)(v_i-u_i)\partial_i T(\rr,t) \nonumber\\ &+& m \Bigl
( (v_i-u_i)(v_k-u_k)- \frac{(\vv-\uu)^2}{3}\delta_{ij}
\Bigr)\partial_i u_k (\rr,t)\Bigr ].
\label{perturbed}
\eea The actual solution $\delta f$, which can be determined by
solving numerically eq. \eqref{unobrey} contains higher-order terms,
but we shall not try to go beyond the approximation \eqref{perturbed}
in the present paper.  Instead, we determine
the kinetic contribution to the transport coefficients.  We first
compute the heat flux vector by substituting $\delta f$ in eq.\eqref{qkin}
\be
q_i(\rr,t)=-\frac{5}{2}\frac{1}{m \nu_0} n(\rr,t) k_B^2 T(\rr,t) \partial_i T(\rr,t)
\label{qexplicit}
\ee
and secondly we compute the components of the pressure tensor \eqref{pkin}
\be
P^{(K)}_{ij}(\rr,t)=k_B T(\rr,t) n(\rr,t)\delta_{ij}
-\frac{1}{\nu_0}\frac{n(\rr,t) k_B T(\rr,t)}{m}\Bigl ( (\partial_i u_j (\rr,t)+\partial_j u_i(\rr,t) )-\frac{2}{3}
\partial_k u_k(\rr,t) \delta_{ij}\Bigr)
\label{pexplicit}
\ee 
By comparing  \eqref{qexplicit} with the macroscopic expression $q_i=-\lambda \partial_i T$ we obtain
the kinetic contribution to the heat conductivity
\be
\lambda^{(K)}=\frac{5}{2}\frac{1}{m \nu_0} n k_B^2 T
\ee
By 
comparing \eqref{pexplicit} with the macroscopic definition of pressure tensor,
\be
P^{(K)}_{ij}(\rr,t)=k_B T n\delta_{ij}-
\Bigl (\eta^{(K)} (\partial_i u_j +\partial_j u_i )+(\eta^{(K)}_b
-\frac{2}{3}\eta^{(K)})
\partial_k u_k \delta_{ij}\Bigr)
\ee
where $\eta_b$ is the bulk viscosity coefficient,
we find the kinetic contribution to the shear viscosity coefficient 
\be
\eta^{(K)}=\frac{n k_B T}{\nu_0}
\label{kshear}
\ee
and
\be
\eta_b^{(K)}=0
\ee
It is important to stress that in the Euler approximation it is not possible to have a stationary solution because
there is no dissipation mechanism to release the energy injected by an external force, so that viscosity and heat conduction
are necessary.
Finally, in order to make contact with the literature one can fix the free parameter of the theory,  $\nu_0$,
and choose $\nu_0=n\sigma^2 \sqrt{2 k_B T/m}$, with $\sigma$ the diameter
of the equivalent hard sphere, and find $\eta^{(K)}=\frac{k_B T m}
{\sqrt 2 }\sigma^2$.

\subsection{Short range repulsive potentials}

To proceed further,  one must solve eq. \eqref{unobrey} and obtain 
explicit expressions for the thermodynamic fields
and the transport coefficients. Thus, we need a specific form of the
interaction potential $U(r,r')$ and consequently of 
$\Omega(\rr,\vv,t)$.
We shall relate the microscopic details to the transport 
coefficients
so that we need to compute the quantities
$C^{(1)}_i$ and $C^{(2)}$.

The prototypical short range repulsive potential is represented by the hard
sphere potential for which one has to consider a special treatment
of the interaction term in order to obtain an accurate representation of the
excess quantities. 
In particular, such an interaction can be treated as a collision process and the collisions as
an uncorrelated binary sequence. A first approximation
of $\Omega[f](\rr_1,\vv,t)$ is given
by the ``Stosszhal ansatz'' which renders \eqref{uno} an equation involving 
only the single-particle distribution decoupled  from higher-order distribution
functions
\bea
\Omega_B[f](\rr,\vv,t) &=& \sigma^{2}\int d\vv_2\int 
d\hat{\ss}\Theta(\hat{\ss}\cdot \vv_{12}) (\hat{\ss} 
\cdot \vv_{12})\times\nonumber\\
&&\left[ f (\rr,\vv',t)f (\rr,\vv_2',t)-f(\rr,\vv,t)f(\rr,\vv_2,t)\right]
\label{collisionb}
\eea
where $\vv'$ and $\vv_2'$ are scattered velocities given by
$\vv'=\vv-(\hat\ss\cdot\vv_{12})\hat\ss$ 
and $\vv_2'=\vv_2+(\hat\ss\cdot\vv_{12})\hat\ss$ with $\vv_{12}=\vv-\vv_2$.

 At higher densities, however, the ``Stosszhal ansatz'' fails to 
describe both the structural and relaxational  properties of the fluid because
collision sequences become highly correlated. To include these sequences and 
extend the transport equation to higher densities in the Seventies van Beijeren
and Ernst have developed the revised Enskog theory (RET), taking into account
the effects of ternary and higher-order collisions and 
the difference in positions of two hard-spheres at collision. Such a feature
allows the instantaneous transfer of momentum and energy during a collision.
In particular this transport mechanism  gives rise to non-ideal
gas contributions to the pressure and to the heat flux, which were
neglected in Boltzmann's treatment of collisions. The RET collision operator
takes the form
\bea
\Omega_{RET}[f](\rr,\vv,t)
&=& \sigma^{2}\int d\vv_2\int 
d\hat{\ss}\Theta(\hat{\ss}\cdot \vv_{12}) (\hat{\ss} 
\cdot \vv_{12})\times\nonumber\\
&&[ g_2(\rr,\rr-\ss\shat,t|n) f (\rr,\vv',t)f (\rr-\ss\shat,\vv_2',t)
\nonumber\\  
&-&
g_2(\rr,\rr+\ss\shat,t|n)f(\rr,\vv,t)f(\rr+\ss\shat,\vv_2,t)]
\label{collision}
\eea

In order to obtain an explicit
representation of the excess 
pressure tensor and heat flux we must compute,
as prescribed by \eqref{divpressurea} and \eqref{divqc}, 
the integrals of
$\Omega_{RET}(\rr,\vv,t)$ times
$G^{(1)}_i(\vv)=m(v-u(\rr,t))_i$
and  $G^{(2)}(\vv)=
\frac{m}{2}(\vv-\uu(\rr,t))^2$.
After
replacing $(\vv,\vv_2,\shat)\rightarrow(\vv',\vv_2',-\shat)$,
the RET explicit form of eqs. (\ref{divpressurea},\ref{divqc}) reads
\bea
&&C_{\underline{\alpha}}^{(l)}(\rr,t) = 
\frac{\sigma^2}{2}\int d\vv\int d\vv_2\int d\hat{\ss}
\Theta(\hat{\ss}\cdot \vv_{12})(\hat{\ss}\cdot \vv_{12})
[G^{(l)}_{\underline{\alpha}}(\vv')-G^{(l)}_{\underline{\alpha}}
(\vv)]\times\nonumber\\
&&
\left[g_{2}(\rr,\rr+\ss\shat,t|n)f(\rr,\vv,t)f(\rr+\ss\shat,\vv_2,t)
-g_{2}(\rr,\rr-\ss\shat,t|n)f(\rr,\vv_2,t)f(\rr-\ss\shat,\vv,t)\right]
\label{cgeneric}
\eea

In the case of the hard-sphere fluid, where there is no contribution
to the internal energy stemming from the pair potential, 
one finds the simple relation 
\be 
C_{HS}^{(2)}(\rr,t)=-[\nabla_i
q_i^{(C)}(\rr,t)+P_{ij}^{(C)}(\rr,t)\nabla_i\uu_j(\rr,t)], 
\ee where
the first term represents the divergence of the heat flux and the
second the contribution due to the viscous heating. 

At this stage we simplify drastically the calculation of the
interaction term and neglect the contribution $\delta f(\rr,\vv,t)$
to the integrals \eqref{cgeneric}. We assume $f(\rr,\vv,t)=
n(\rr,t)\phi_M(\rr,\vv,t)$, which depends on the three hydrodynamic
fields $n,u,T$. Substituting now this approximation into 
\eqref{cgeneric} and expanding to first order
$\phi_M(\rr\pm\shat,\vv,t)$
about $\uu(\rr,t)$ and $T(\rr,t)$
\bea
\phi_M(\rr\pm\ss\shat,\vv,t) \simeq  \left[\frac{m}{2\pi k_B T_(\rr,t)}\right]^{3/2}\exp
\left[-\frac{m(\vv-\uu(\rr,t))^2}{2 k_B T(\rr,t)} \right]\times\nonumber\\
\big\{ 1+\frac{m[\vv-\uu(\rr,t)]\cdot[\uu(\rr\pm\ss\shat)-
\uu(\rr)]}{k_B T(\rr,t)}\nonumber\\
+\frac{m[\vv-\uu(\rr,t)]\cdot
[\vv-\uu(\rr,t))-3 k_B T(\rr,t)]}
{2 k_B T^2(\rr,t)} [T(\rr\pm\ss\shat,t)-T(\rr,t)] +...\big\}
\nonumber 
\eea
we arrive, after some lengthy algebra, to the result

\bea
&&C^{(1)}_i(\rr,t) = 
-k_B T(\rr,t) n(\rr,t)\sigma^2
\int d\hat{\ss}\hat{\ss}_i
g_{2}(\rr,\rr+\ss\shat,t|n)n(\rr+\ss\shat,t)
\nonumber\\
&&\bigg[1-
\frac{2}{\sqrt{\pi k_B T(\rr,t)/m}} \shat\cdot [\uu(\rr+\ss\shat,t)-\uu(\rr,t)]
+\frac{T(\rr+\ss\shat,t)-T(\rr,t)}{2 T(\rr,t)} \bigg]
\label{c1expl}
\eea
and

\bea
&&C^{(2)}(\rr,t) = {k_B T(\rr,t)} n(\rr,t)\sigma^2
\int d\hat{\ss} g_{2}(\rr,\rr+\ss\shat,t|n) n(\rr+\ss\shat,t)   
\nonumber\\
&&\bigg[
-\frac{\shat\cdot[\uu(\rr+\ss\shat,t)-\uu(\rr,t)]}{2}
+  \frac{1}{\sqrt\pi}\sqrt{\frac{k_B T(\rr,t)}{m}}
\frac{T(\rr+\ss\shat,t)-T(\rr,t)}{T(\rr,t)} \bigg]
\label{cc2c}
\eea

A comment is in order. 
In the hard-sphere fluid, momentum and energy
fluxes can be transferred instantaneously even when
the velocity distribution function
has a Maxwellian form, provided it peaks at
the local hydrodynamic velocity with a
spread determined by the local temperature.
As a consequence, we obtain a contribution not only to the 
pressure but also
to the transport
coefficients even within a Maxwellian approximation to the
distribution function. This result is at variance with 
the corresponding result
in the case of the Boltzmann equation where the collision term
does not contribute neither to the pressure nor to the transport coefficients 
when the distribution is Maxwellian.

\subsection{Collisional contribution to the transport coefficients in the Hard-Sphere fluid}

We now apply our method to compute the fluid transport coefficients in bulk conditions.

{\it Shear viscosity.}
We assume both the density and the temperature
to be uniform throughout  the system, i.e. $n(\rr,t)=n_0$ and $T(\rr,t)=T_0$
whereas the velocity profiles varies along a direction normal
to the stream lines
\be
\uu=(0,u_y(x,0),0)
\ee
We then use the equations
\be
P_{xy}=P_{yx}=
-\eta[\frac{\partial u_y}{\partial x}+\frac{\partial u_x}{\partial y}]
\label{comp1}
\ee
and
\be
C^{(1)}_y=-\frac{\partial P_{xy}^{(C)}}{\partial x}=
\eta^{(C)} \frac{\partial^2 u_y}{\partial x^2}
\label{comp}
\ee
in order to establish a relation between the macroscopic
$\eta$ coefficient and the microscopic level represented by $C^{(1)}_i$.

Using eq. \eqref{c1expl}
and expanding to second order in $\sigma$ we obtain
\be
 C_y^{(1)}(\rr,t)= mv_T^2\sigma^{4}g_{2}(\sigma)n_0^2
\frac{1}{\sqrt{\pi}v_T} \frac{\partial^2 u_{y}}{\partial x^2}
\int d^{d+1}\hat{\sigma}
\hat{\sigma}_{y}^2\hat{\sigma}_{x}^2=
mv_T \sigma^4 g_{2}(\sigma)n_0^2
\frac{1}{\sqrt{\pi}} \frac{\partial^2 u_{y}}{\partial x^2}
\frac{4\pi}{15}
\label{cc1}
\ee
where $g_2(\sigma)$ is the radial distribution function at contact.
By comparing \eqref{cc1} and \eqref{comp} we obtain
\be
\eta^{(C)}=\frac{4}{15} m v_T \sqrt\pi \sigma^4 g_{2}(\sigma) n_0^2
=0.266\sqrt{k_B T \pi m}g_{2}(\sigma) n_0^2 \sigma^4
\ee
which is the viscosity found by Longuet-Higgins and Pople  \cite{Longuet} the case
of hard-spheres,
whereas Enskog formula \cite{SiguHeyes}  at high density gives
$\eta^{(C)}_{Enskog }=0.337\sqrt{k_B T \pi m}g_{2}(\sigma) n_0^2 \sigma^4$.

{\it Bulk viscosity.}
Similarly if one assumes $\uu=(u_x(x,0),0,0)$, recalls
$C_x^{(1)}(\rr,t)=-\frac{\partial P_{xx}}{\partial x}$
and uses the macroscopic relation 
$$\frac{\partial P_{xx}}{\partial x}
=-(\eta_b+\frac{4}{3}\eta)\frac{\partial^2 u_x}{\partial x^2}
$$
can extract the bulk viscosity.
Expanding $C_x^{(1)}$ to second order in $\sigma$ we find
\be
 C_x^{(1)}(\rr,t)= mv_T^2\sigma^{4}g_{2}(\sigma)n_0^2
\frac{1}{\sqrt{\pi}v_T} \frac{\partial^2 u_{x}}{\partial x^2}
\int d^{4}\hat{\sigma}
\hat{\sigma}_{x}^4=
mv_T \sigma^4 g_{2}(\sigma)n_0^2
\frac{1}{\sqrt{\pi}} \frac{\partial^2 u_{x}}{\partial x^2}
\frac{4\pi}{5}
\ee
That is $\eta_b^{(C)}= \frac{5}{3}\eta^{(C)}$.

{\it Thermal conductivity.}
The thermal conductivity can be computed 
assuming that both the density and the velocity
are uniform throughout  the system, i.e. $n(\rr,t)=n_0$ and $\uu(\rr,t)=0$.
In this case the heat flux is proportional to the temperature
gradient: ${\bf q^{(C)}}(\rr,t)=-\lambda^{(C)} \nabla T(\rr,t)$
and $C^{(2)}(\rr,t)=-\nabla\cdot {\bf q^{(C)}}(\rr,t)$.
We assume a temperature profile varying only along the x-direction
$T(x,t)$ and use eq.\eqref{cc2c}, so that after some simple 
calculations and expanding to second order in $\sigma$ we obtain
\be
C^{(2)}(\rr,t)= m v_T^3 g_{2}(\sigma)n_0^2 \sigma^{4}
\frac{1}{\sqrt{\pi}}\frac{k_B}{m v_T^2}
\frac{1}{2}\nabla^2 T(\rr,t)\int d^{d-1}\hat{\sigma} \hat{\sigma}_{x}^2 
=m v_T \sigma^{4} g_{2}(\sigma)n_0^2
\frac{k_B}{m }
\frac{2\sqrt\pi}{3}\nabla^2 T(\rr,t) 
\ee

The heat conductivity HS excess is
\be
\lambda^{(C)}=\frac{2 }{3}\sqrt{k_B T \pi m}  g_2(\sigma) n_0^2 \sigma^{4}
\frac{k_B}{m }=0.666 \sqrt{k_B T \pi m}  g_2(\sigma) n_0^2 \sigma^{4}
\frac{k_B}{m }
\ee


while Enskog formula  at high density gives
$\lambda^{(C)}_{Enskog}=1.269 \sqrt{k_B T \pi m}g_{2}(\sigma) n_0^2 \sigma^4
\frac{k_B}{m }$


\subsection{Beyond hard core dynamics}

In the case of more realistic potentials, such as the Lennard-Jones
interaction, a satisfactory theory for dense systems is still missing.
The situation has been reviewed recently by Stell and coworkers 
\cite{Stell2007}.
They compared different theories of transport
in the Lennard-Jones fluid. These include a kinetic variational theory
(KVT)  \cite{Stell1985}, and
a stochastic approach put forward in its original form
by Leegwater \cite{Leegwater} and later reformulated by Polewczak and Stell
\cite{Stell2002} which
introduces a  random distribution of diameters of hard-spheres in the collision term. 

The KVT applies to Lennard-Jones (LJ) fluids and is obtained by
adding a hard-sphere core to the attractive tail of the LJ potential. The transport coefficients exhibit Enskog-like forms, but the radial distribution function bears explicit dependence on the LJ tail as well as on the hard-sphere core. The hard sphere diameter is determined according to the well-known WCA method used in equilibrium statistical mechanics to mimic the LJ fluid \cite{WCA}. 

We shall not try to apply these theories in the present
paper, a task beyond our scope, although 
the stochastic method could perhaps be implemented in our scheme.
We shall only comment that the difficulty for continuous interactions
is that there are not instantaneous collisions followed by free streaming
trajectories as in the case of hard spheres. Instead, the collisions have a 
finite duration and a second collision event can take place before
the  first one is completed.
A possible workaround could be to consider  the motion of
each particle  as the combination of continuous momentum changes
and  almost instantaneous collisions. These small momentum 
changes represent the effect of  weak attractive forces exerted by the surrounding
molecules and can be assimilated to the random white noise forces
occurring in the Brownian motion.  In the literature, Rice and Allnatt 
combined the hard-core and the soft fluctuating potential model \cite{Kreuzer,RiceAllnatt}.
They represented the collisional
term as a sum of the Enskog-Boltzmann collision integral, describing
the large  momentum binary exchanges   plus
a Fokker-Planck collision operator,
supplemented by an average force, to model the small momentum exchanges 
due to the attractive forces,
and assumed that these  processes proceed without mutually interfering.

This separation should be feasible, since
in many simple fluids the pair potential can be decomposed into the sum
of a short range harsh repulsive potential contribution, plus
a an attractive long range tail. The fine details of such a 
decomposition depend on the accuracy of the approximation, the most popular
methods being the Weeks-Chandler-Andersen (WCA) and the Barker-Henderson \cite{BarkerHenderson} ones.
It is therefore possible to treat the short range part with the technique
employed in the case of the hard spheres and the tail of the interaction
in the adiabatic  approximation illustrated
in section \ref{colloid}.  On the other hand, there are 
other possibilities.
One of them consists in approximating the full continuous potential 
by a stepwise
potential for which one can derive a proper extension of the RET equation, 
but the method seems  exceedingly
cumbersome to be of practical relevance.

In conclusion, as far as we are interested in studying systems of
particles with realistic potentials we can treat the attractive
tails in a mean field (Vlasov) fashion and the repulsion as if determined by a
hard-spheres with a diameter suitably chosen.
Since   $C_i^{(1)}$ and $ C^{(2)}$ do not have dependence on velocity and 
temperature gradients, they  cannot contribute to the 
transport coefficients.
To be more explicit, by projecting the interaction $\Omega$ over the
hydrodynamic space,
the resulting ``matrix elements'' do not depend on the fields
$\uu$ and $T$, i.e. the Vlasov approximation only accounts
for the distortion of the density $n$ from its equilibrium value. Thus,
the transport coefficients are only related to the 
BGK term and are purely kinetic within this approximation.


\section{Solution by Lattice Boltzmann strategy}
\label{lbesolution}
The Lattice Boltzmann represents an alternative to methods which involve
the direct solutions of coupled hydrodynamic equations of the type
\eqref{continuity}, \eqref{momentumcontinuity} and \eqref{energyequation}.
These equations represent projections of the evolution, eq.\eqref{unobrey}, and are
not closed, since even the kinetic contributions of the pressure tensor
and heat flux vector are not expressible solely in terms of
the $n$, ${\bf u}$ and $T$. The necessary information to determine
these contributions is contained in the higher moments of the distribution
function. One can avoid the difficulty of computing these moments
by using a macroscopic point of view which assumes constitutive relations,
i.e. introduces phenomenological linear relations between $P_{ij}$
and ${\bf q}$ and the velocity and temperature, respectively. Alternatively,
one is faced with the problem of closing a dynamical infinite hierarchy
of equations. We instead propose to solve eq.\eqref{unobrey} directly,
with the only approximation stemming from the discretization procedure.

The LB method first evolved as empirical extension of lattice gas
automata and found wide application as a standard simulation tool
in computational fluid-dynamics \cite{LBgeneral}. In subsequent years, the LB method
was found to be a systematic procedure to solve numerically
a kinetic equation in velocity space \cite{Shanhe,Martysshanchen,Heluo}.
Having in mind the simulation of condensed systems in micro/mesoscopic
conditions, the critical parameter governing the departure from equilibrium
is the Knudsen number, $\epsilon$, being the ratio between the mean
free path and the representative length scale. On the nanoscale, $\epsilon$
can be arbitrarily large and the numerical method should offer great
flexibility with respect to external conditions and the Knudsen value. 

As previously shown, a direct solution of eq. \eqref{uno} is numerically very demanding 
in the case of a realistic form of the collision term $\Omega(f)$ due to the
large number of integrations over the phase-space variables.
In this section we focus on the general strategy to solve the kinetic
equation \eqref{unobrey} in the adiabatic approximation, i.e. for $C^{(2)}=0$
and the system is rigorously isothermal. However, extending these
concepts to the presence of heat transport is rather straightforward,
along the lines discussed below. Moreover, we assume that the functional
$g_{2}(\rr,\rr',t|n)$ is completely known by some level of theory or from
atomistic simulations, such as Molecular Dynamics.
In the study of fluids under non-equilibrium conditions, the proposed scheme 
has a strategic advantage over conventional atomistic simulations schemes since it
exploits the pre-averaged nature of the kinetic description, 
avoiding the need to average observables over different realizations of the noise.

The starting point to derive a consistent numerical scheme is the
representation for both the distribution function and the collisional
terms over a finite set of Hermite polynomials. The distribution $f(\rr,\vv,t)$
is approximated by
\begin{equation}
\bar{f}(\rr,\vv,t)=\omega(v)\sum_{l=0}^{K}\frac{1}{v_{T}^{2l}2l!}\phi^{(l)}(\rr,t)h^{(l)}(\vv)
\end{equation}
with $\omega(v)=(2\pi v_{T}^{2})^{-3/2}e^{-v^{2}/2v_{T}^{2}}$.
To keep the notation compact, in the following
we use the convention that the product of tensors implicitly indicates
the sum over all permutations of the tensorial indices. By construction,
the complete and truncated distributions have the same coefficients
up to Hermite order $K$, i.e.
\begin{equation}
\int d\vv h^{(l)}(\vv)f(\rr,\vv,t)=
\int d\vv h^{(l)}(\vv)\bar{f}(\rr,\vv,t)\,\,\,\,\,\,\,\,\,\,\mbox{for}\,\,\, l\leq K
\end{equation}
Being the distribution coefficients $\phi^{(l)}$ a combination of
the distribution moments up to $l$ order, the full and truncated
distributions share the same moments up to $K$ order. 

Similarly, the collision operator is replaced by a truncated representation
over the same Hermite set. We distinguish between the non-hydrodynamic relaxation
term, taken here as the BGK relaxation \eqref{simple}, 
and the collisional term, renamed as 
${\cal K}(\rr,\vv,t) \equiv 
\frac{1}{m} C_{i}^{(1)}(\rr,\vv,t) \partial_{v_{i}} \Phi_M(\rr,\vv,t) \equiv
\frac{1}{mn(\rr,\vv,t)} C_{i}^{(1)}(\rr,\vv,t) \partial_{v_{i}}f^{eq}(\rr,\vv,t)$.
Alternatively, the straight heat-bath operator (\ref{liquidheatbath}) or 
momentum-preserving version (\ref{liquidheatbath}) can be chosen and
treated along similar lines \cite{FP,FPH}. In this way, the colloidal/density functional dynamics 
or a different microscopic dynamics for the non-hydrodynamic modes can be selected.
The corresponding truncated representations are
\be
\overline{\delta\Omega}(\rr,\vv,t)=\omega(v)\sum_{l=0}^{K}\frac{1}{v_{T}^{2l}2l!}\psi^{(l)}(\rr,t)h^{(l)}(\vv)
\ee
and
\be
\overline{{\cal K}}(\rr,\vv,t)=\omega(v)\sum_{l=0}^{K}\frac{1}{v_{T}^{2l}2l!}{\cal \chi}^{(l)}(\rr,t)h^{(l)}(\vv)
\ee
Again, each collisional term shares the same $K$ moments with the
original collisional counterpart.

The next step is to employ Gauss-Hermite quadratures to evaluate hydrodynamic
and collisional moments. Recognizing that $\bar{f}(\rr,\vv,t)h^{(l)}(\vv)/\omega(v)=p(\rr,\vv,t)$
is a polynomial in $\vv$ of degree $\leq2K$, the moments can
be evaluated exactly with quadratures of order $2G\geq K$, since
\begin{eqnarray}
\phi^{(l)}(\rr,t) & = & \int d\vv\bar{f}(\rr,\vv,t)h^{(l)}(\vv)=\int d\vv\omega(\vv)p(\rr,\vv,t)\nonumber \\
 & = & \sum_{p=0}^{G}w_{p}p(\rr,{\bf c}_{p},t)=\sum_{p=0}^{G}f_{p}(\rr,t)h^{(l)}({\bf c}_{p})\end{eqnarray}
where $f_{p}(\rr,t)\equiv w_{p}\bar{f}(\rr,{\bf c}_{p},t)/\omega(c_{p})$,
the vectors ${\bf c}_{p}$ are a set of quadratures nodes and $w_{p}$ the associated
quadrature weights. Similarly, the collisional moments are computed
as 
\begin{eqnarray}
\psi^{(l)}(\rr,t) & = & \sum_{p=0}^{G}\delta\Omega_{p}(\rr,t)h^{(l)}({\bf c}_{p})\\
\chi^{(l)}(\rr,t) & = & \sum_{p=0}^{G}{\cal K}_{p}(\rr,t)h^{(l)}({\bf c}_{p})
\end{eqnarray}

In summary, with the truncated Hermite representation and Gauss-Hermite
quadratures, the distribution is replaced by an array of $Q$ populations,
$f(\rr,\vv,t)\rightarrow f_{p}(\rr,t)=w_{p}\bar{f}(\rr,{\bf c}_{p},t)/\omega(c_{p})$
and similarly for the collisional term, $\delta\Omega(\rr,\vv,t)\rightarrow\delta\Omega_{p}(\rr,t)=w_{p}\overline{\delta\Omega}(\rr,{\bf c}_{p},t)/\omega(c_{p})$
and ${\cal K}(\rr,\vv,t)\rightarrow{\cal K}_{p}(\rr,t)=w_{p}\bar{{\cal K}}(\rr,{\bf c}_{p},t)/\omega(c_{p})$.
The nodes ${\bf c}_{p}$ are chosen as vectors connecting neighboring
mesh points $\rr$ on a lattice, mirroring the hop of particles
between mesh points, generally augmented by a null vector ${\bf c}_{0}$
accounting for particles at rest. The specific form of the lattice
velocities and weights depends on the order of accuracy of the method
and reflects the required Hermite order, as described in the following
and thoroughly discussed in ref. \cite{shanyuangchen}.
The mesh is a Cartesian grid and
the lattice velocities satisfy the sum rules, $\sum_{p}w_{p}c_{pi}=0$,
$\sum_{p}w_{p}c_{pi}c_{pj}=v_{T}^{2}\delta_{ij}$, $\sum_{p}w_{p}c_{pi}c_{pj}c_{pk}=0$
and $\sum_{p}w_{p}c_{pi}c_{pj}c_{pk}c_{pl}=v_{T}^{4}(\delta_{ik}\delta_{jl}+\delta_{il}\delta_{jk})$,
in order to guarantee mass and momentum conservation and isotropy. 

The third step of the procedure is to propagate the distribution via
a discretization of the streaming operator to first order, as a forward
Euler update, 
\begin{equation}
f_{p}(\rr+\Delta t{\bf c}_{p},t+\Delta t)=
f_{p}(\rr,t)+\Delta t w_{p}\sum_{l=0}^{K}\frac{1}{v_{T}^{2l}l!}\left[\psi^{(l)}(\rr,t)+\chi^{(l)}(\rr,t)\right]h^{(l)}({\bf c}_{p})
\end{equation}
where $\Delta t$ is the LB time-step.
Consequently, the algorithm exploits the same Cartesian mesh to rearrange populations
over spatial and velocity shifts.

The population dynamics is able to reproduce the target macroscopic
evolutions, such as the Navier-Stokes equation, to high accuracy.
In the appendix a formal Chapman-Enskog multiscale analysis
shows that for a second-order of accuracy 
in the transport coefficients, a second order expansion in Hermite is
needed for both the equilibrium distribution and the collisional integral.
As a result, the final form of the LB algorithm reads
\begin{equation}
f_{p}(\rr+{\bf c}_{p},t+1)=(1-\frac{\Delta t}{\tau})f_{p}(\rr,t)+\frac{\Delta t}{\tau}f_i^{eq}(\rr,t)+\Delta t {\cal K}_i(\rr,t)
\end{equation}
where the relaxation time $\tau$ is related to the kinetic component of the shear viscosity via 
\begin{equation}
\eta^{(K)}=nv_{T}^{2}(\tau-\frac{\Delta t}{2})
\label{visctau}
\end{equation}
i.e. the physical value (compare with  eq. \eqref{kshear})  minus a viscosity of numerical origin. 
Some straightforward algebra shows that the final form of the discrete equilibrium reads
\begin{equation}
f_{p}^{eq}=w_{p}n(\rr,t)\left[1+\frac{c_{pi}u_{i}(\rr,t)}{v_{T}^{2}}+\frac{(c_{pi}c_{pj}-v_{T}^{2}\delta_{ij})u_{i}(\rr,t)u_{j}(\rr,t)}{2v_{T}^{4}}\right]\end{equation}
which is tantamount to a low-Mach ($O[Ma^{3}]$) expansion of the
local Maxwellian, while the discretized form of the collisional term is
\begin{equation}
{\cal K}_{p}=-w_{p}\frac{1}{m}\left[\frac{c_{pi}C_{i}^{(1)}(\rr,t)}{v_{T}^{2}}+\frac{(c_{pi}c_{pj}-v_{T}^{2}\delta_{ij})u_{i}(\rr,t)C_{j}^{(1)}(\rr,t)}{v_{T}^{4}}\right]
\end{equation}
A popular mesh model that reproduces the second order Hermite accuracy is provided by
the so-called D3Q19 model \cite{LBgeneral}, consisting of $19$ discrete velocities in three
dimensions, and with $v_{T}=1/\sqrt{3}$.

A further issue concerns the calculation of the spatial convolution
present in the collisional term. From the above discussion it is clear
that the central issue in LB related methods is the discretization
of velocity space with the ensuing level of accuracy in the macroscopic
transport equations. On the other hand, the method is completely flexible
in terms of the mesh spacing $\Delta x$, that can be tuned at will in order to
resolve the details of the microscopic interactions. Therefore, the
error introduced in the spatial discretization does not represent
a critical issue. As an example, in a previous paper \cite{Melchionna2008}
we solved numerically eq. \eqref{unobrey} by means of the Lattice
Boltzmann method \cite{LBgeneral} using as a test case the flow of
a hard-sphere fluid through a narrow channel with a pressure gradient
along its axis, and measuring the deviations from the macroscopic
Poiseuille law. The spatial convolutions were evaluated
with spatial quadratures over a number of off-mesh points obtained
via linear interpolations, one of several alternatives to compute
such integrals to desired level of accuracy.

Regarding the stability of the proposed method, it is worth mentioning that even
for simple non-interacting dynamics the LB algorithm is subjected to numerical
instability. By using a von Neumann linear stability analysis, it
has been shown that a stable LB scheme requires that the flow velocity
be below a certain threshold that is a function of the relaxation
time and the wave number \cite{sterlingchen}. The action of stiff
intermolecular forces clearly narrows the stability range.
Nevertheless, a generic upper bound for the variation of
populations due to the forcing term is $\delta f/f\sim f_{p}/w_{p}\sim\Delta t{\cal K}_{p}\ll1$.
Therefore,
for a generic quadrature scheme, the convolution force is $C^{(1)}/\Delta x^{3}\sim nFg_{2}$,
and the forcing term should be $nFg_{2}\ll1/\Delta x^{4}$
since $\Delta x/\Delta t\sim1$, showing that the stability upper
bound raises rapidly when reducing the mesh spacing.


\section{Conclusions}
\label{conclusions}
To summarize, we have presented a theoretical analysis and 
proposed a computational scheme
which bridge hydrodynamics with microscopic structural theories of fluids.
The present approach shows that the dynamic density functional and Boltzmann-like methods can 
be derived in a unique framework. The differences between the two methods 
are determined by the interaction of the fluid with the heat bath.
The DDF method applies to colloids which are embedded in a solvent
whose degrees of freedom are represented by a viscous heat bath, not moving
with the particles, which
eliminates all the hydrodynamic modes, but the diffusive mode.
Boltzmann methods, instead, apply to molecular fluids 
and the corresponding heat bath
is determined by degrees of freedom internal to the fluids and therefore
moves with the fluid. 

A further improvement could concern the dependence of the  hydrodynamic fields
$\uu(\rr,t)$ and $T(\rr,t)$ on the local distribution $f(\rr,\vv,t)$. 
One could expect that these fields 
should be obtained  via a  coarse-grained  prescription 
starting from $\phi_M(\rr,\vv,t)$
because  the hydrodynamics cannot be
extended beyond the size of few diameters.
In fact, at scales shorter than a few molecular diameters hydrodynamic modes 
are strongly damped, with a decay time so short that the time
autocorrelation function of a single molecule is isotropic. Therefore,
descriptions based on
hydrodynamic field could become inappropriate at that scale \cite{Delgado}.
This program has been carried on for the hard sphere model 
in ref. \cite{Melchionna2008}  by employing
a radial distribution function 
constructed according to the prescription of
Fischer and Methfessel \cite{fishmet}. It is assumed that
$g_2 (\rr,\rr'|n)$ depends on a a
coarse-grained density $\bar n (\rr,t)$ via a uniform smearing over a
sphere of radius $\sigma/2$. As a consequence the collisional terms  
\eqref{c1expl} and \eqref{cc2c} depend non locally not only from the density as in DDF but also from the
temperature and velocity field.

In the final section of the paper we showed how the derived kinetic
equations can be transformed into a numerical scheme by following
the theoretical derivation, i.e. by relying on a truncated Hermite
representation of the collisional integrals and the unknown
distribution, complemented by Gauss-Hermite quadratures to evaluate
the distribution moments.  Previous implementations of these ideas by
us have proved that this numerical approach can be successfully applied
to the study of inhomogeneous fluids.

\section{Acknowledgments}
UMBM thanks Pedro Tarazona for correspondence.

\appendix
\section{Chapman-Enskog analysis}

Let us analyze the multi-scale dynamics in powers of the Knudsen number $\epsilon$,
with the distribution expanded as\begin{equation}
f=f^{eq}+\epsilon f^{(1)}+\epsilon^{2}f^{(2)}+...\end{equation}
The dynamics is further decomposed based on the intrinsic hierarchy of timescales.
For example, one can distinguish between the
short-time convective motion, over a spatial scale $x\rightarrow\epsilon x$
and temporal scale $t_{1}=\epsilon t$, and the slower evolution set by momentum
diffusion, $t_{2}=\epsilon^{2}t$. 
Expressing the derivatives as
\begin{eqnarray}
\partial_{t} & = & \epsilon\partial_{t}^{(1)}+\epsilon^{2}\partial_{t}^{(2)}+...\\
\partial & = & \epsilon\partial
\end{eqnarray}
the Lattice Boltzmann streaming step is expressed as 
\begin{eqnarray*}
f_{p}(\rr+{\bf c}_{p},t+1)-f_{p}(\rr,t) & = & (\epsilon\partial_{t}^{(1)}+\epsilon^{2}\partial_{t}^{(2)}+...+\epsilon c_{p}\partial)(f_{p}^{eq}+\epsilon f_{p}^{(1)}+...)
\end{eqnarray*}
having set $\Delta t=1$.
We now employ the explicit form of the kinetic equation \eqref{unobrey}. The
collision operator has an intrinsic dependence on $\epsilon$, and
the BGK term reads
\begin{equation}
\frac{1}{\tau}(f_{p}^{eq}-f_{p})=-\frac{1}{\tau}(\epsilon f_{p}^{(1)}+\epsilon^{2}f_{p}^{(2)}+...)
\end{equation}
where, as shown next, the relaxational time $\tau$ is related to
the kinetic component of the shear viscosity. The collisional term
in \eqref{unobrey} is a non-linear functional of the density, multiplied by
$\partial_{v}f^{eq}$, that depends on density and momentum density,
so that the fastest evolution is on the $\epsilon$ scale, \begin{equation}
{\cal K}=\epsilon{\cal K}^{(1)}+\epsilon^{2}{\cal K}^{(2)}+...\end{equation}
and ${\cal K}^{(0)}=0$. By equating the same orders in $\epsilon$,
the evolutions on the $\epsilon$ scale is
\begin{eqnarray}
(\partial_{t}^{(1)}+c_{p}\partial)f_{p}^{eq} & = & -\frac{f_{p}^{(1)}}{\tau}+{\cal K}_{p}^{(1)}
\label{1stce}
\end{eqnarray}
and on $\epsilon^{2}$ scale
\begin{eqnarray}
&&\partial_{t}^{(1)}f_{p}^{(1)}+\partial_{t}^{(2)}f_{p}^{eq}+\partial c_{p}f_{p}^{(1)}+\frac{1}{2}c_{p}c_{p}\partial^{2}f_{p}^{eq}+\partial\partial_{t}^{(1)}c_{p}f_{p}^{eq}+\frac{1}{2}\partial_{t}^{(1)}\partial_{t}^{(1)}f_{p}^{eq} 
\nonumber\\
& = & -\frac{f_{p}^{(2)}}{\tau}+{\cal K}_{p}^{(2)}
\label{2ndce0}
\end{eqnarray}
Substituting eq.(\ref{1stce}) into the fourth and last terms in the l.h.s. of eq.(\ref{2ndce0}), 
and by rearranging terms, the $\epsilon^{2}$ dynamics can be rewritten as
\begin{eqnarray}
\left(\partial_{t}^{(1)}+c_{p}\partial\right)\left[\left(1-\frac{1}{2\tau}\right)f_{p}^{(1)}+\frac{1}{2}{\cal K}_{p}^{(1)}\right]+\partial_{t}^{(2)}f_{p}^{eq} & = & -\frac{f_{p}^{(2)}}{\tau}+{\cal K}_{p}^{(2)}
\label{2ndce}
\end{eqnarray}

Up to the momentum diffusivity scale, therefore, the evolution of
the density and current are obtained by contracting the above equations
as $\sum_{p}(\cdot)$ and $\sum_{p}{\bf c}_{p}(\cdot)$, obtaining
\begin{eqnarray}
\partial_{t}^{(1)}n+\partial_{i}^{(1)}(nu_{i}) & = & 0\label{eq:denseps}\\
\partial_{t}^{(2)}n & = & 0\label{eq:denseps2}\\
\partial_{t}^{(1)}(nu_{i})+\partial_{j}\left(nv_{T}^{2}\delta_{ij}+nu_{i}u_{j}+P_{ij}^{(C,1)}\right) & = & 0\label{eq:curreps}\\
\partial_{t}^{(2)}(nu_{i})+\frac{1}{\epsilon}\partial_{j}\left(\left(1-\frac{1}{2\tau}\right)P_{ij}^{(K,neq)}+P_{ij}^{(C,2)}\right) & = & 0\label{eq:curreps2}
\end{eqnarray}
where $P_{ij}^{(C,1)}$ and $P_{ij}^{(C,2)}$ are the collisional
contributions to the pressure tensor at first and second Knudsen order,
respectively. In addition, some lengthy algebra shows that $P_{ij}^{(K,neq)}$
is identified as the non-equilibrium part of the kinetic pressure
tensor 
\begin{equation}
P_{ij}^{(K,neq)}=\left(1-\frac{1}{2\tau}\right)\sum_{p}c_{pi}c_{pj}(f_{p}-f_{p}^{eq})=\eta^{(K)}[\partial_{i}u_{j}+\partial_{j}u_{i}]
\end{equation}

By multiplying eqs. \eqref{eq:denseps} and \eqref{eq:denseps2} by
$\epsilon$ and eqs. \eqref{eq:curreps} and \eqref{eq:curreps2}
by $\epsilon^{2}$ and summing the equations, we reconstruct the sought
evolution for density and momentum density accurate to $\epsilon^{2}$
level 
\begin{eqnarray*}
\partial_{t}n+\partial_{i}(nu_{i}) & = & 0\\
\partial_{t}(nu_{i})+\partial_{j}\left(nv_{T}^{2}\delta_{ij}+P_{ij}^{(K)}+P_{ij}^{(C)}\right) & = & 0
\end{eqnarray*}

At higher Knudsen order, the following recurrence
relation holds \cite{shanyuangchen}
\begin{equation}
f_{p}^{(k+1)}=-\tau\left[\sum_{m=0}^{k}\partial_{t}^{(k)}f_{p}^{(m)}+c_{p}\partial f_{p}^{(k)}-{\cal K}_{p}^{(k+1)}\right]
\end{equation}
where $f^{(0)}\equiv f^{eq}$. Similarly, for the Hermite coefficients,
\begin{equation}
\phi^{(n,k+1)}=-\tau\left[\sum_{m=0}^{k}\partial_{t}^{(k)}\phi^{(n,m)}+n\partial\phi^{(n-1,k)}+\partial\phi^{(n+1,k)}-\chi^{(n,k+1)}\right]\label{eq:pyramid}
\end{equation}
where the first superscript of the coefficients refers to the Hermite
truncation level and the second to the Knudsen level. If we neglect
the collisional term, eq. \eqref{eq:pyramid} shows that the dynamics
of the coefficient $\phi^{(n,k+1)}$ depends on $\phi^{(n-1,k)}$,
$\phi^{(n+1,k)}$ and $\phi^{(n,0)}$, ..., $\phi^{(n,k)}$ in a hierarchical
(pyramidal) way. In order to guarantee a given level of accuracy on
the Knudsen level, without introducing any approximation from the
Hermite truncation, a complete representation is required on the base
of the pyramid corresponding to the equilibrium distribution. In particular,
for Knudsen order $k$, the equilibrium needs to include at least
$n+k+1$ Hermite coefficients \cite{shanyuangchen}. At first order
in Knudsen, the equilibrium distribution should be accurate up to
third Hermite order, complemented by fifth order quadratures to evaluate
the moments. In order to handle thermal dynamics (in the presence
of a non-zero heat flux), a fourth order accuracy in the equilibrium
is required, together with seventh order quadratures. 

When the collisional term is included, the simple hierarchical picture
seems to be spoiled. However, by using the explicit Hermite representation
of the collisional term, the third order Hermite coefficient $\chi^{(3)}\sim u^{3}$,
a contribution that can be neglected in standard condensed matter
conditions. If one retains only up to second order in $f_{p}^{(eq)}$,
the error can be estimated, as discussed in \cite{shanyuangchen},
to be $\sim\frac{1}{\nu_{0}}u^{2}\partial_{x}u$ in the pressure tensor
and $\sim u^{2}$ in the shear viscosity. 



\end{document}